\documentclass[aps,prb,twocolumn,showpacs,superscriptaddress,groupedaddress]{revtex4-1}
\usepackage{dsfont,amsthm,amsbsy}
\usepackage{amssymb}
\usepackage{amsmath}
\usepackage{bbm}
\usepackage{graphicx}
\usepackage{epstopdf}
\usepackage{subfigure}
\usepackage{natbib}
\usepackage{epsfig}
\usepackage{amsfonts}
\usepackage{mathrsfs}
\usepackage{sidecap}
\usepackage{lipsum}
\usepackage[toc,page,title,titletoc,header]{appendix}
\usepackage[colorlinks,linkcolor=blue,citecolor=blue,anchorcolor=blue]{hyperref}
\hypersetup{ hidelinks, }
\usepackage{dsfont,amsthm,amsbsy}
\usepackage{resizegather}
\usepackage{tikz}
\usepackage{float}
\newcommand{\be}{\begin{equation}}
\newcommand{\ee}{\end{equation}}
\newcommand{\bea}{\begin{eqnarray}}
\newcommand{\eea}{\end{eqnarray}}

\newcommand{\bi}{\begin{itemize}}
\newcommand{\ei}{\end{itemize}}
\newcommand{\bc}{\begin{center}}
\newcommand{\ec}{\end{center}}

\begin{document}

\title{Multicriticality in a one-dimensional topological band insulator}

\author{Mariana Malard$^{1,2}$, David Brandao$^{1}$, Paulo Eduardo de Brito$^{1}$, and Henrik Johannesson$^{2}$}
\affiliation{$\mbox{}^1$Faculty of Planaltina, University of Brasilia, 70904-910, Brasilia-DF, Brazil}
\affiliation{$\mbox{}^2$Department of Physics, University of Gothenburg, SE 412 96 Gothenburg, Sweden}

\begin{abstract}

A central tenet in the theory of quantum phase transitions (QPTs) is that a nonanalyticity in a ground-state energy implies a QPT. Here we report on a finding that challenges this assertion. As a case study we take a phase diagram of a one-dimensional band insulator with spin-orbit coupled electrons, supporting trivial and topological gapped phases separated by intersecting critical surfaces. The intersections define multicritical lines across which the ground-state energy becomes nonanalytical, concurrent with a closing of the band gap, but with no phase transition taking place.

\end{abstract}


\maketitle

\section{Introduction}

Phase transitions are ubiquitous in all branches of physics. Present-day research has largely focused on equilibrium transitions at zero temperature $-$ {\em quantum phase transitions} (QPTs) $-$ identified by the appearance of a nonanalyticity in the ground-state energy of a system in the thermodynamic limit \cite{Sachdev}. A QPT is called \emph{first-order} when the first derivative of the ground-state energy with respect to a control parameter (a coupling constant or an external field) becomes discontinuous, or {\em continuous} when the discontinuity appears in the second or a higher derivative. These definitions mimic their classical counterparts from equilibrium thermodynamic phase transitions described in terms of a nonanalytical canonical free energy \cite{Kastner}.

The conventional picture of continuous QPTs (excluding infinite-order QPTs of Berezinskii-Kosterlitz-Thouless type \cite{B,KT}) is that they entail a change of symmetry or topology of a ground state. Symmetry-breaking QPTs are described by the Landau-Ginzburg-Wilson paradigm \cite{ChaikinLubensky} or that of ``deconfined quantum criticality" \cite{Senthil}, being key to a number of theories of quantum matter, implying universal scaling behavior of observables at low temperatures \cite{Sachdev,ContinentinoBook}. Topological QPTs in turn can be broadly classified into transitions between symmetry-protected \cite{Chiu2016} or topologically ordered \cite{Wen} phases, characterized by a topological invariant or ground state degeneracy.

In both symmetry-breaking and topological QPTs, the competition between two incompatible ground states is thought to be the cause of the nonanalyticity in the ground-state energy \cite{ContinentinoBook}. Here we report on a finding which calls for a reexamination of this assumption. Studying a model of non-interacting spin-orbit coupled electrons on a one-dimensional (1D) lattice, we uncover a phase diagram consisting of topologically nontrivial and trivial gapped phases separated by multiple intersecting gapless (critical) surfaces. The crossing between two such surfaces defines a multicritical line with an unexpected property: The ground-state energy develops a nonanalyticity in the thermodynamic limit, accompanied by a closing of the energy gap to the first excited level, suggestive of a continuous QPT. Yet there is no change of symmetry or topology across the multicritical line. While a nonanalytic behavior without a spontaneous symmetry breaking or change of topology $-$ but with scaling of the type seen in a continuous QPT $-$ has been conjectured to become possible at certain putative ``quantum-critical end points" (see Ref. \onlinecite{KirkpatrickBelitz} and references therein), our result is obtained by an exact analysis of a well-defined Hamiltonian. By this, the phenomenon of a nonanalyticity without a QPT is put on firm ground, but now at quantum multicriticality in a topological phase diagram.

\section{Model}

We consider a 1D lattice with $N$ sites populated by electrons with nearest-neighbor hopping and subject to Dresselhaus and Rashba spin-orbit interactions, the latter being spatially modulated. The corresponding tight-binding Hamiltonian writes:
\begin{equation}\label{H}
H\,=\,\sum^{N}_{n=1}\sum_{\alpha,\alpha' = \uparrow, \downarrow}\,h_{\alpha\alpha'}(n)\,c^{\dag}_{n,\alpha}\,c_{n+1,\alpha'}\,+\,\mbox{H.c.},
\end{equation}
where $c^{\dag}_{n,\alpha}$ ($c_{n,\alpha}$) is the creation (annihilation) operator for an electron at site $n$ with spin projection ${\alpha}\!=\,\uparrow,\downarrow$ along a $z$-quantization axis. The matrix elements are given by $h_{\alpha\alpha'}(n)\,=\,-t\delta_{\alpha\alpha'}-i\gamma_{\text{D}}\sigma_{\alpha\alpha'}^{x}-i\gamma_{\text{R}}(n)\sigma_{\alpha\alpha'}^{y}$, with $\sigma^{x(y)}$ the $x$\,($y$) Pauli matrix and the real parameters $t$, $\gamma_{\text{D}}$, and $\gamma_{\text{R}}(n)$ being the amplitudes of hopping, Dresselhaus, and Rashba spin-orbit coupling respectively. With this choice of basis, the chain is along the $x$-axis. The Rashba parameter is spatially modulated as $\gamma_{\text{R}}(n) = \gamma_{\text{R}}+\gamma'_{\text{R}}\cos(2\pi qn+\phi)$, with $2\pi q/a$ being the wave number ($a$ is the lattice spacing) and $\phi$ the phase of the modulation with respect to the underlying lattice. This Hamiltonian belongs to the class of generalized Aubry-Andr\'e-Harper models {\cite{Harper,Aubry}. In Ref. \onlinecite{Ortix2015} a similar Hamiltonian was derived from an effective description of a curved quantum wire. A different realization, with an added periodic chemical potential and electron interactions, may be obtained by gating a quantum wire with an array of nano-sized electrodes \cite{Malard2011}.
\begin{figure*}[htpb]
  \includegraphics[width=14cm]{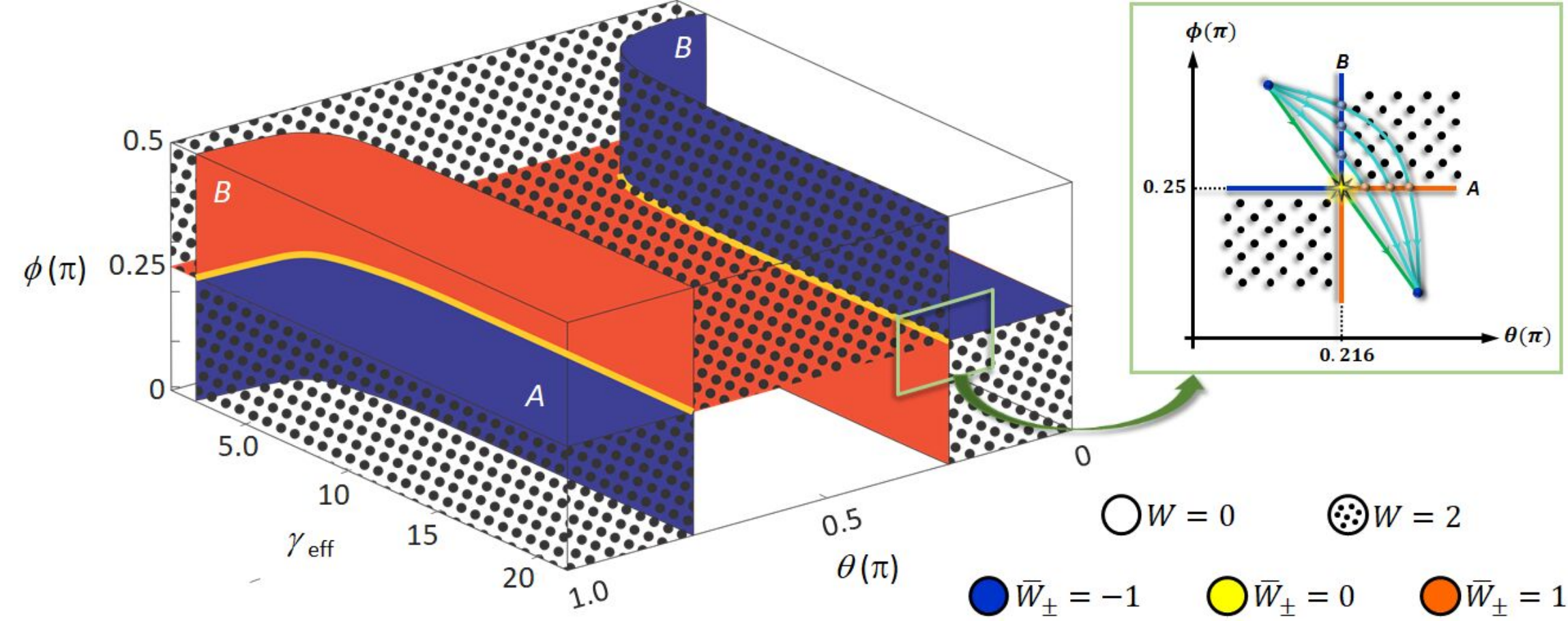}
  \caption{(Color online) Winding number $W$ and local winding numbers $\bar{W}_{\pm}$ in the three-dimensional $(\gamma_{\text{eff}},\theta,\phi)$ parameter space of the model. The phase diagram consists of topologically nontrivial insulating phases (dotted) where $W=2$ and trivial insulating phases (empty) where $W=0$, separated by critical surfaces $A$ and $B$ where the spectrum displays a pair of gap-closing nodes symmetrically located in the Brillouin zone with $\bar{W}_{\pm}=-1$ (blue), $\bar{W}_{\pm}=1$ (orange), and $\bar{W}_{\pm}=0$ (yellow), with the latter defining the multicritical lines of the model. In the inset, paths that connect two regions with $W=0$ are shown on a large-$\gamma_{\text{eff}}$ cross section of the phase diagram. Along the curved paths the system undergoes two consecutive second-order topological QPTs, the first at the critical line $B$ and the second at $A$. As the paths approach the limiting straight line that crosses the multicritical point, the second-order topological QPTs become closer in parameter space and eventually merge into a fourth-order nonanalyticity at the multicritical point.}
\end{figure*}

With periodic boundary conditions, $H$ is invariant under translations by a
unit cell, each cell containing $r=1/q$ sites. A Fourier transform yields the single-particle Bloch Hamiltonian represented by a $2r\times2r$ matrix ${\cal H}(k) = adiag(Q(k),Q^{\dagger}(k))$, with $Q(k)$ an $r\times r$ matrix. The formalism is detailed in Appendix A. Choosing, for example, $r=4$,
\begin{equation} \label{Qk}
Q(k)\,=\,\begin{bmatrix}
    A_{1} & e^{-ik}\!A^{\ast}_{4}\\
    A^{\ast}_{2} & \ \ \ \ A_{3}  \\
\end{bmatrix},
\end{equation}
where $A_{n}$, $n=1,...,4$, are the $2\times2$ matrices
\begin{equation} \label{A}
A_{n}\,=\,\begin{bmatrix}
    \alpha_{n}^{+} & \beta_{n}\\
    \beta_{n} & \alpha_{n}^{-}
\end{bmatrix}
\end{equation}
with diagonal [off-diagonal] entries given by spin-conserving [spin-flipping] hopping amplitudes. With $t=1$ and $\gamma'_R=13.5$, one finds that $\alpha_{n}^{\pm}=-1\mp i\gamma_{\text{eff}} \mp i\cos(\theta)\cos(\pi n/2+\phi)$ and $\beta_{n}=i\sin(\theta)\cos(\pi n/2+\phi),$ with $\gamma_{\text{eff}}=\sqrt{\gamma^{2}_{R}+\gamma^{2}_{D}}$ and $\theta = \arctan(\gamma_D/\gamma_R)$. By this, the model is fully parametrized by $\gamma_{\text{eff}}, \theta$, and $\phi$.

When $r$ is an even integer, the $2r\times2r$ Bloch Hamiltonian ${\cal H}(k)$ belongs to symmetry class CII of the Altland-Zirnbauer classification \cite{Chiu2016}, being invariant under chiral and time-reversal symmetry (see Appendix B). Differently, the two-band version of the model, with $r=1$ site per unit cell, supports only time-reversal symmetry and, thus, belongs to class AII which is trivial in 1D \cite{Chiu2016}. Generating a non-trivial topology from a trivial two-band model by increasing the number of bands is an interesting possibility surfaced by the present model.

\section{Topological Phase Diagram}

The gapped phases of the half-filled $2r$-band model in (\ref{H}) (with even $r$) are distinguished by the $2Z$ winding number of class CII \cite{Chiu2016}, call it $W$. For the points in parameter space where the band gap closes, the half-filled gapless spectrum can be characterized by `local' winding numbers $\bar{W}_{\pm}$ \cite{Li2015} (with $\pm$ here labelling two gap-closing points symmetrically located in the Brillouin zone (BZ)), analogous to how a Weyl node in a semimetal is characterized by a topological charge \cite{Armitage}. For details, see Appendix C.

Fig. 1 shows the result of a numerical computation of $W$ and $\bar{W}_{\pm}$ in the three-dimensional $(\gamma_{\text{eff}},\theta,\phi)$ parameter space when $r=4$, $t=1$ and $\gamma'_{R}=13.5$. We constrain $\theta\in[0,\pi]$ and $\phi\in[0,\pi/2]$ since the phase diagram is periodic with period $\pi$ along $\theta$ and $\pi/2$ along $\phi$. The phase diagram consists of topologically nontrivial [trivial] gapped phases $-$ the dotted [empty] regions where $W=2$ [$W=0$] $-$ separated by critical surfaces colored in orange [blue] if the corresponding gap-closing nodes carry local winding numbers $\bar{W}_{\pm}=1$ [$\bar{W}_{\pm}=-1$]. The intersections of the critical surfaces, depicted in yellow, define multicritical lines along which $\bar{W}_{\pm}=0$.
\begin{figure}
\includegraphics[scale=0.540]{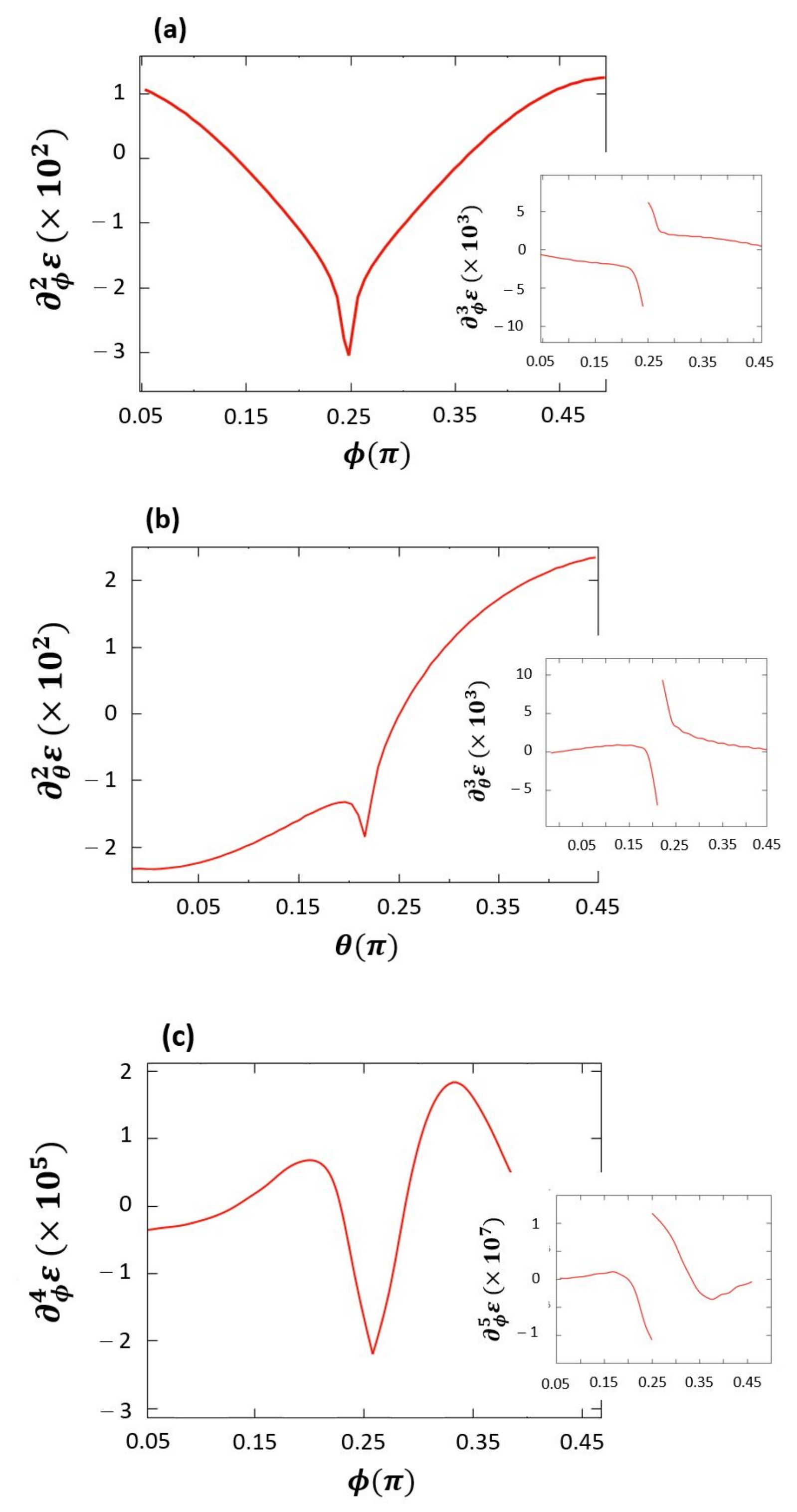}
\caption{(Color online) (a), (b): Second-order derivative of the ground state energy density $\varepsilon$ for a topological QPT between a trivial $W\!=\!0$ phase and a topologically nontrivial $W\!=\!2$ phase across (a) the critical plane $A$ with $\theta=0.1\pi$ and (b) the critical surface $B$ with $\phi=0.1\pi$. (c): Fourth-order derivative of $\varepsilon$ for a $W\,=\,0\,\rightarrow\,0$ path parametrized by $\theta=-(0.216/0.25)(\phi-0.5\pi)$ which cuts through the multicritical line. The insets in (a)-(c) show the next derivative of $\varepsilon$. In all panels, $\gamma_{\text{eff}}=20$ (cf. Fig. 1). The numerical data was obtained for a chain with $10^{4}$ unit cells for which $\varepsilon$ has converged to the thermodynamic limit and has become independent of system size (see Figs. 6 and 7 in Appendix E).}
\end{figure}

The critical surfaces come in two types: the plane $A$ at $\phi=\pi/4$, and the surfaces $B$ which are curved towards $\theta=0,\pi$ for small $\gamma_{\text{eff}}$ and become flat for large $\gamma_{\text{eff}}$. The reason for choosing $r=4$ can now be explained: This choice provides the minimal realization of the considered class of models supporting a multicritical phase diagram. Indeed, with $r=4$ sites per unit cell, the model acquires an off-centered mirror symmetry when $\phi=\pi/4$ and this symmetry (additional to the time-reversal and chiral symmetries of class CII) forces the band gap to close at zero energy (cf. Appendix D). Multicritical lines are thus generated from the crossings of the plane $A$ (defined by $\phi=\pi/4$) and the curved critical surfaces $B$ of the model. We note that while $A$ implies symmetry-enforced band touchings in the BZ, the band touchings associated with the surfaces $B$ are accidental, with no extra symmetry enforcing them. The closing of the gap at $A$ and $B$ happens through the formation of a pair of time-reversal symmetric zero-energy nodes defining the apexes of two 1D Dirac cones in the BZ and carrying equal local winding numbers $\bar{W}_{+}=\bar{W}_{-}$. On the $A$-surface, and hence also on the the multicritical lines, these winding numbers serve as topological invariants which code for the symmetry enforcement.

{\em Multicriticality} $-$ Having resolved the phase diagram, we come to our main result: the anomalous critical behavior across the multicritical lines in Fig. 1.

We first check the character of the topological QPTs across $A$ - Fig. 2(a) - and $B$ - Fig. 2(b) - when going between a trivial $W\!=\!0$ phase and a topological $W\!=\!2$ phase: In both cases the second derivatives of the ground-state energy density $\varepsilon$ (defined as the ground state energy divided by the number of unit cells) of a half-filled chain develops a cusp - as attested by the discontinuous third derivative shown in the insets - at the critical value of the control parameter $\theta$ or $\phi$, signaling a continuous QPT.

By crossing a multicritical line along a path which connects two regions in the phase diagram with the {\em same} $W$, one obtains a different result. Now the derivatives of the ground-state energy are smooth up to third order (cf. Appendix E), but with a cusp developing in the fourth derivative when hitting the multicritical line; cf. Fig. 2(c) for a $W=0\rightarrow0$ transition along the path $\theta = -(0.216/0.25)(\phi-0.5\pi)$ depicted in the inset of Fig. 1. Mirroring this path in the $\phi=0.25\pi$ line in the inset of Fig. 1 realizes a $W\!=\!2\rightarrow\!2$ transition, producing a mirror image of the plot in Fig. 2(c). Importantly, the cusp in the fourth derivative is insensitive to the choice of path through the multicritical line.

Fig. 3 shows the band gap $\Delta$ on the $\theta\times\phi$ plane for $\gamma_{\text{eff}}=20$, with the values of the winding number $W$ of the gapped phases indicated on the gap surface. The gap displays a conical profile close to the topological QPTs connecting phases with different values of $W$. In contrast, the fourth-order nonanalyticity across a multicritical line linking phases with equal $W$ correlates with a parabolic gap-closing, as obtained from analyzing a cross section of Fig. 3 along the path defined in Fig. 2(c).

In a phase with $W=2$, the critical behavior comes with a divergence in the localization length $\xi$ of the topological edge states. For transitions away from [through] a multicritical point, $\xi$ scales linearly [parabolically] with the inverse distance to the critical point \cite{unpublished}, implying linear scaling between $\Delta$ and $\xi$ and a dynamical critical exponent $z=1$. These results confirm predictions from a recent formulation of scaling laws in topological QPTs \cite{Chen4}, suggesting that it may also be useful for describing topological phase diagrams where criticality does not coincide with a change of a topological invariant.
\begin{figure}
\includegraphics[scale=0.35]{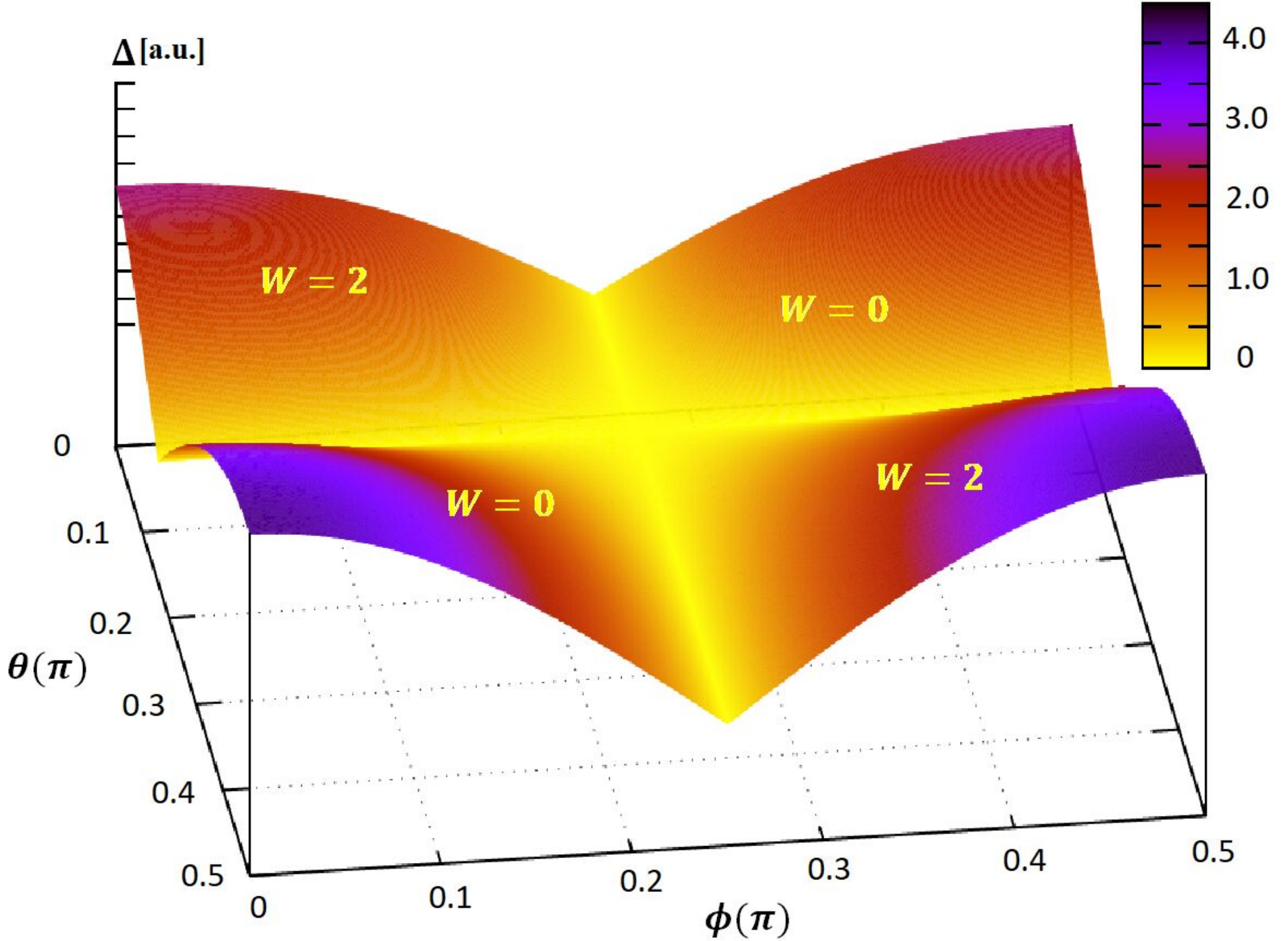}
\caption{(Color online) Band gap $\Delta$ on the $\theta\times\phi$ plane when $\gamma_{\text{eff}}=20$ (with $\Delta$ and $\gamma_{\text{eff}}$ in arbitrary units [a.u.]). The winding number $W$ of the gapped phases is indicated on the gap surface. The figure indicates that the closing of the gap is conical for the second-order topological QPTs connecting phases with different $W$. Taking a cross section of the above gap surface along any path through the multicritical point (e.g. the path defined in Fig. 2(c)) parametrizing a fourth-order $W\,=\,0\,\rightarrow\,0$ transition, and fitting the data, we obtain a parabolic gap-closing: $\Delta=179|\phi-0.25|^{z\nu}$ to the left and $\Delta=112|\phi-0.25|^{z\nu}$ to the right of the multicritical point, with the gap critical exponent \cite{Sachdev} $z\nu=2.00\pm0.07$.}
\end{figure}

\section{Discussion}

In the conventional theory of continuous QPTs, a nonanalyticity in the ground-state energy does not happen except at a critical point separating phases which are distinguished either by symmetry or by topology \cite{Sachdev,ContinentinoBook}. The present simple model of a topological band insulator defies this notion: The ground-state energy becomes nonanalytical at multicritical lines in the phase diagram, with associated closing of the band gap and divergence of a localization length, and yet there is no change of symmetry or in the topological invariant. A spontaneous symmetry breaking is excluded by the fact that an insulating ground state $-$ whether in a topologically trivial or nontrivial phase $-$ is unique when periodic boundary conditions are imposed, with the many-particle ground state formed by a Slater determinant of the single-particle states of the filled bands. With the two regions carrying the same winding numbers $W$, there is clearly no change of topology as defined by the Altland-Zirnbauer (AZ) classification \cite{Chiu2016}. Therefore, if conventional theory is strictly true, and having excluded $-$ on general grounds $-$ a change of symmetry, it follows that the equal-$W$ phases connected by the multicritical line must be distinguishable by a new topological invariant beyond the AZ scheme \cite{Chiu2016}.

As a case in point, it is well known that by adding space group symmetries to the AZ scheme produces a richer classification of possible topological phases \cite{Ando}. In one spatial dimension, from the possible space group symmetries - translation by a unit cell, inversion, and mirror (1D inversion $\times$ a spin flip in a spinful system) - only mirror symmetry can lead to new topological phases beyond the AZ classification \cite{Chiu2016}. While our 1D model has a special off-centered mirror symmetry on its gapless plane $A$ (see Fig. 1 and Appendix D), there is no mirror symmetry in the gapped phases and, thus, they are exhaustively classified by the AZ scheme, at least as far as space group symmetries go. The source of a topological distinction would therefore have to be something else. Notably, a hypothetical new topological invariant \emph{can not} be tied to a bulk-boundary correspondence since the insulators with $W=0$ have no in-gap boundary states. This makes the existence of such an invariant seem unlikely.

Still in the context of a hypothetical topological transition beyond the AZ scheme, the nonanalyticity at the multicritical line could be the result of a Lifshitz transition. Such a transition entails a change of the Fermi-surface topology, with a jump of its Euler characteristic \cite{Volovik}. Note, however, that a Fermi surface of a 1D electron system, as in our case study, consists of disconnected spin-degenerate points. A change of the topology requires an explicit symmetry breaking, producing an energy mismatch between the Fermi points for spin-up and spin-down populations $-$ such as in the well-known magnetic Lifshitz transitions in 1D; see e.g. Ref. \onlinecite{Ptok} and references therein. This scenario does not play out in the present case, and thus the transition across the multicritical line is not a Lifshitz transition.

Having excluded symmetry-breaking and topological transitions, we conjecture that a ground state in the {\em proximity} to a QPT between symmetry-protected topological phases may occasionally develop a critical behavior {\em without} undergoing a QPT. Note that we are here concerned only with true ``in-phase" critical points. Phase diagrams in which a critical point connecting equal phases is a result of a redundancy is clearly of no interest; here one of the phases can be folded on top of the other via a gauge transformation \cite{MBBJnote}. For the type of phase diagram as in Fig. 1, one may interpret the occurrence of a nonanalyticity at the multicritical line as a ``remnant" of the normal nonanalyticities at the two intersecting critical surfaces. But how do the two second-order nonanalyticities then conspire to produce one at higher order, falsely signaling a QPT?

The conventional theory of multicritical behavior in symmetry-breaking QPTs is not of much help. Such transitions are driven by quantum fluctuations in a local order parameter, governed by the competing renormalization-group (RG) fixed points for the critical lines that meet at multicriticality \cite{Boettcher}. This produces a multicritical behavior with new critical and crossover exponents, or the appearance of a first-order transition,  but not one of higher order \cite{FisherBook}. As for QPTs of topological band insulators, as in the present case, the transitions are not caused by disordering fluctuations in a local order parameter, but instead by a rearrangement of the phase structure of the single-particle Bloch states that build the ground state. This opens up for phenomena not encountered in symmetry-breaking QPTs, the unexpected interplay between topology and multicriticality uncovered in our work being an example. While our finding has been established for a particular class of Hamiltonians, similar results can be derived for other models exhibiting topological multicriticality \cite{InProgress}; examples include the Haldane model for a Chern insulator \cite{Haldane}, the Creutz model with induced superconductivity \cite{Sticlet}, and the dimerized Kitaev chain \cite{Wakatsuki}.

It is an open problem to explore the full implications of our finding, maybe taking off from recent attempts to formulate an RG approach to QPTs between symmetry-protected topological phases \cite{Continentino1,Continentino2,RoyGoswamiSau,Chen1,Chen2,Chen3,Griffith,Niewenburg,Chen4,RLCG,Continentino3}.

\section*{Acknowledgements}

We thank Wei Chen, Carlos Egues, Gia Japaridze, Cristiane Morais Smith, Andreas Schnyder, and Stellan \"Ostlund for valuable discussions. This work was supported by the Swedish Research Council through Grant No. 621-2014-5972.\\ \\

\section*{Appendix}

\subsection{Bloch Hamiltonian}

In this appendix the Bloch Hamiltonian presented in Eqs. (\ref{Qk}) and (\ref{A})
is derived. For that, let us reintroduce the original tight-binding model in position space in Eq. (1):
\begin{equation} \label{SH}
H\,=\,\sum^{N}_{n=1}\sum_{\alpha,\alpha' = \uparrow, \downarrow}\,h_{\alpha\alpha'}(n)\,c^{\dag}_{n,\alpha}\,c_{n+1,\alpha'}\,+\,\mbox{H.c.},
\end{equation}
where
\begin{equation}\label{h}
h_{\alpha\alpha'}(n)\,=\,-t\delta_{\alpha\alpha'}-i\gamma_{\text{D}}\sigma_{\alpha\alpha'}^{x}-i\gamma_{\text{R}}(n)\sigma_{\alpha\alpha'}^{y},
\end{equation}
with $c^{\dag}_{n,\alpha}$ ($c_{n,\alpha}$) the creation (annihilation) operator for an electron at site $n$ with spin projection ${\alpha}\!=\,\uparrow,\downarrow$ along a $z$-quantization axis, $\sigma^{x(y)}$ the $x$\,($y$) Pauli matrix, and the real parameters $t$, $\gamma_{\text{D}}$, and $\gamma_{\text{R}}(n)=\gamma_{\text{R}}+\gamma'_{\text{R}}\cos(2\pi qn+\phi)$ the amplitudes of hopping, Dresselhaus spin-orbit coupling, and the spatially modulated Rashba spin-orbit coupling respectively; $2\pi q/a$ is the wave number ($a$ being the lattice spacing) and $\phi$ is the phase of the modulation.

Performing a rotation of basis that diagonalizes the uniform part of $H$ in spin space:
\begin{align}
\nonumber &d_{n,+}=\frac{1}{\sqrt{2}}(e^{-i\theta/2}c_{n,\uparrow}-ie^{i\theta/2}c_{n,\downarrow}),\\
\nonumber &d_{n,-}=\frac{1}{\sqrt{2}}(-ie^{-i\theta/2}c_{n,\uparrow}+e^{i\theta/2}c_{n,\downarrow}),
\end{align}
with $\theta=\arctan(\gamma_{\text{D}}/\gamma_{\text{R}})$, Eqs. (\ref{SH})-(\ref{h}) take on the form
\begin{equation} \label{Hrotated}
H\,=\,\sum^{N}_{n=1}\sum_{\tau =\pm}\,[\alpha_{n}^{\tau}\,d^{\dag}_{n,\tau}\,d_{n+1,\tau}\,+\,\beta_{n}\,d^{\dag}_{n,\tau}\,d_{n+1,-\tau}]\,+\,\mbox{H.c.},
\end{equation}
where $\tau=\pm$ labels the spin projections along the new quantization axis determined by the combination of Dresselhaus and Rashba couplings and where the strength $\alpha_{n}^{\tau}$ of the spin-conserving and $\beta_{n}$ of the spin-flipping hopping are given by
\begin{eqnarray}
\label{alpha} \alpha_{n}^{\tau}\,&=&\,-(t+i\tau\gamma_{\text{eff}})-i\tau\gamma'_{\text{R}}\cos(\theta)\cos(2\pi qn+\phi), \\
\label{beta} \beta_{n}\,&=&\,i\gamma'_{\text{R}}\sin(\theta)\cos(2\pi qn+\phi),
\end{eqnarray}
where $\gamma_{\text{eff}} = \sqrt{\gamma_R^2 + \gamma_D^2}$.

Imposing periodic boundary conditions, $H$ is translation invariant on a lattice with $M=N/r$ unit cells with $r=1/q$
sites per unit cell. Eq. (\ref{Hrotated}) can thus be rewritten in terms of intra-cell and inter-cell contributions as
\begin{equation} \label{Hintrainter}
H\,=\,\sum^{M}_{m=1}\left[\sum^{r-1}_{n=1}\sum_{\tau=\pm}{\cal H}_{\text{intra}}\,+\,\sum_{\tau=\pm}{\cal H}_{\text{inter}}\right]\,+\,\mbox{H.c.},
\end{equation}
with
\begin{equation} \label{calHintra}
{\cal H}_{\text{intra}}\,=\,\alpha_{n}^{\tau}\,d_{m,n}^{\tau\,\dagger}d_{m,n+1}^{\tau}\,+\,\beta_{n}\,d_{m,n}^{\tau\,\dagger}d_{m,n+1}^{-\tau},
\end{equation}
\begin{equation} \label{calHinter}
{\cal H}_{\text{inter}}\,=\,\alpha_{r}^{\tau}\,d_{m,r}^{\tau\,\dagger}d_{m+1,1}^{\tau}\,+\,\beta_{r}\,d_{m,r}^{\tau\,\dagger}d_{m+1,1}^{-\tau},
\end{equation}
where $d_{m,n}^{\tau\,\dagger}$ ($d_{m,n}^{\tau}$) creates (annihilates) a particle at site $n$ in unit cell $m$ with spin projection $\tau$. Fig. 4 illustrates the spin-conserving and the spin-flipping hoppings within and across a unit cell.
\begin{figure}
\includegraphics[scale=0.3]{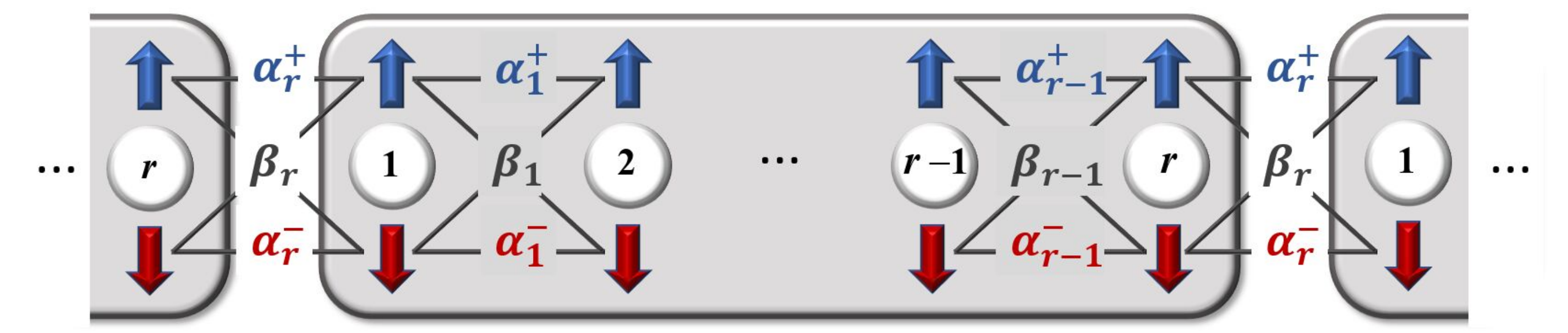}
\caption{(Color online) Intra(Inter)-cell hoppings acting inside (across) a unit cell of $r$ sites; $\alpha$-hoppings are spin-conserving and $\beta$-hoppings are spin-flipping.}
\end{figure}

By Fourier transforming the electron operators with respect to the unit cell position coordinate $m$,
\begin{equation} \label{FourierTransform}
d_{m,n}^{\tau}\,=\,\frac{1}{\sqrt{M}}\sum^{\pi}_{k=-\pi}\,d_{k,n}^{\tau}e^{ikm},
\end{equation}
where $k=k_{j}=\pm2\pi j/M$, $j=0,1,...,M/2$, Eqs. (\ref{Hintrainter})-(\ref{calHinter}) turn into
\begin{equation} \label{Hfouriertransformed}
H\,=\,\sum^{\pi}_{k=-\pi}\,\sum^{r}_{n,n'=1}\,\sum_{\tau,\tau'=\pm}\,d_{k,n}^{\tau\dagger}\,{\cal H}_{n\tau,n'\tau'}(k)\,d_{k,n'}^{\tau'}
\end{equation}
with
\begin{eqnarray} \label{MatrixElementsofH}
\nonumber {\cal H}_{n\tau,n'\tau'}(k)&\!=\!&\alpha_{n}^{\tau}\delta_{n',n+1}\delta_{\tau',\tau}\!+\!\beta_{n}\delta_{n',n+1}\delta_{\tau',-\tau}\\
\nonumber &\!+\!&\alpha_{n-1}^{\tau\,\ast}\delta_{n',n-1}\delta_{\tau',\tau}\!+\!\beta_{n-1}^{\ast}\delta_{n',n-1}\delta_{\tau',-\tau} \\
\nonumber &\!+\!&\alpha_{r}^{\tau\,\ast}e^{-ik}\delta_{n,1}\delta_{n',r}\delta_{\tau',\tau}\!+\!\beta_{r}^{\ast}e^{-ik}\delta_{n,1}\delta_{n',r}\delta_{\tau',-\tau} \\
\nonumber &\!+\!&\alpha_{r}^{\tau}e^{+ik}\delta_{n,r}\delta_{n',1}\delta_{\tau',\tau}\!+\!\beta_{r}e^{+ik}\delta_{n,r}\delta_{n',1}\delta_{\tau',-\tau}.
\end{eqnarray}

We proceed by partitioning the lattice into two sublattices - one formed out from the odd-labelled intra-cell sites and the other from the even-labelled ones - and defining a $2r$-dimensional row spinor $d^{\dagger}_{k}$ which groups the creation operators in the following way:
\begin{equation} \label{ChiralSpinor}
d^{\dagger}_{k}\,=\,(d_{k,1}^{+\,\dagger},d_{k,1}^{-\,\dagger},...,d_{k,r-1}^{+\,\dagger},d_{k,r-1}^{-\,\dagger},d_{k,2}^{+\,\dagger},d_{k,2}^{-\,\dagger},...,d_{k,r}^{+\,\dagger},d_{k,r}^{-\,\dagger}),
\end{equation}
where the first (last) $r$ entries receive the creation operators defined on the sublattice of odd (even) intra-cell sites, with operators for up and down spins at the same site placed next to each other. A $2r$-dimensional column spinor $d_{k}$ is defined by grouping the annihilation operators in the same way.

In matrix form the Hamiltonian (\ref{Hfouriertransformed}) thus becomes
\begin{equation} \label{FourierTransformed}
H=\sum_{k=-\pi}^{\pi}\,d^{\dag}_{k}\,{\cal H}(k)\,d_{k},
\end{equation}
where, in the chosen spinor representation (\ref{ChiralSpinor}),
\begin{equation} \label{HkAlgebraic}
{\cal H}(k)\,=\,\frac{\sigma_{x}}{2}\otimes[Q(k)+Q^{\dagger}(k)]\,+\,\frac{i\sigma_{y}}{2}\otimes[Q(k)-Q^{\dagger}(k)]
\end{equation}
or, equivalently,
\begin{equation} \label{Hk}
{\cal H}(k)\,=\, \begin{bmatrix}
    0 & Q(k) \\
    Q^{\dagger}(k) & 0 \\
\end{bmatrix},
\end{equation}
with the $r\times r$ matrix $Q(k)$ given by
\begin{equation} \label{SQk}
Q(k)\,=\, \begin{bmatrix}
    A_{1} & 0 & 0 & \dots & 0 & zA^{\ast}_{r}\\
    A^{\ast}_{2} & A_{3} & 0 & \dots & 0 & 0 \\
    \vdots & \vdots & \vdots & \dots & \vdots & \vdots \\
    0 & 0 & 0 & \dots & A^{\ast}_{r-2} & A_{r-1} \\
\end{bmatrix},
\end{equation}
where $z=e^{-ik}$ and
\begin{equation} \label{SA}
A_{n}\,=\, \begin{bmatrix}
    \alpha_{n}^{+} & \beta_{n}\\
    \beta_{n} & \alpha_{n}^{-}
\end{bmatrix}
\end{equation}
are $2\times 2$ matrices whose diagonal (off-diagonal) entries are given by the spin-conserving (-flipping) hopping amplitudes of Eqs. (\ref{alpha})-(\ref{beta}). For a chain with $r=4$ sites per unit cell, $Q(k)$ is thus given by Eq. (2).

\subsection{Symmetry class}

A single-particle Bloch Hamiltonian ${\cal H}(k)$ is invariant under chiral ($S$), time reversal ($T$), and particle-hole ($C$) transformations if it satisfies the following invariance relations:
\begin{equation} \label{ChiralSymmetry}
{\cal S}\,{\cal H}(k)\,{\cal S}^{-1}\,=\,-{\cal H}(k),
\end{equation}
\begin{equation} \label{TimeReversalSymmetry}
{\cal T}\,{\cal H}(k)\,{\cal T}^{-1}\,=\,{\cal H}^{\ast}(-k),
\end{equation}
\begin{equation} \label{ParticleHoleSymmetry}
{\cal C}\,{\cal H}(k)\,{\cal C}^{-1}\,=\,-{\cal H}^{\ast}(-k),
\end{equation}
where ${\cal S}$, ${\cal T}$ and ${\cal C}$ are matrices representing $S$ and the unitary parts of $T$ and $C$ respectively \cite{Chiu2016}.

Here we define the chiral transformation as ${\cal S}={\cal P}_{\text{o}}-{\cal P}_{\text{e}}$, where ${\cal P}_{\text{o}}$ (${\cal P}_{\text{e}}$) is the projector onto the sublattice of odd (even) intra-cell sites. Given this definition, ${\cal S}$ takes the form ${\cal S}\,=\,\sigma_{z}\otimes1\!\!1_{r\times r}$ in the spinor representation (\ref{ChiralSpinor}). One immediately verifies that the model has $S$-symmetry, i.e. Eq. (\ref{ChiralSymmetry}) is satisfied with ${\cal H}(k)$ given by Eq. (\ref{HkAlgebraic}).

Going to $T$-symmetry, ${\cal T}$ in  Eq. (\ref{TimeReversalSymmetry}) is the matrix performing a spin flip which, in the representation (\ref{ChiralSpinor}), reads ${\cal T}\,=\,1\!\!1_{r\times r}\otimes(-i\sigma_{y})$, with ${\cal T}^{2}\,=\,-1\!\!1$, as required for a spinful electron. The symmetry relation (\ref{TimeReversalSymmetry}) applied to Eqs. (\ref{Hk})-(\ref{SQk}) thus implies: $(-i\sigma_{y})A_{n}(i\sigma_{y})\,=\,A^{\ast}_{n}$, which is indeed satisfied with $A_{n}$ given by Eq. (\ref{A}) and Eqs. (\ref{alpha})-(\ref{beta}).

As for $C$-symmetry, ${\cal S}\,=\,{\cal T}{\cal C}$ implies that ${\cal C}\,=\,-{\cal T}{\cal S}$. Having fulfilled equalities (\ref{ChiralSymmetry})-(\ref{TimeReversalSymmetry}), one checks that Eq. (\ref{ParticleHoleSymmetry}) is then also satisfied. With the above expressions for ${\cal S}$ and ${\cal T}$, we have that ${\cal C}\,=\,\sigma_{z}\otimes1\!\!1_{r/2\times r/2}\otimes(i\sigma_{y})$, and hence ${\cal C}^{2}\,=\,-1\!\!1$.

The above results for ${\cal S}^{2}$, ${\cal T}^{2}$ and ${\cal C}^{2}$ puts our $2r$-band realization of the Hamiltonian (\ref{FourierTransformed}) with $r\in2\mathds{Z}$ sites per unit cell in symmetry class CII of the Altland-Zirnbauer classification, with its gapped phases being labeled by a $2\mathds{Z}$-winding number \cite{Chiu2016}. By the bulk-boundary correspondence, this implies that each edge of the chain hosts two zero-energy states, implying that the zero-energy level is four-fold degenerate. This is easy to verify numerically by computing eigenenergies and wave functions for the Hamiltonian in Eqs. (\ref{SH})-(\ref{h}) using open boundary conditions \cite{unpublished}.

\subsection{Winding numbers}

{\em Gapped phases.} The $2\mathds{Z}$-winding number of the gapped phases, denoted by $W$, is defined as the number of times that the complex number $\text{det}[Q(k)]$, with $Q(k)$ given in Eqs. (\ref{SQk})-(\ref{SA}), winds around the origin of the complex plane as $k$ sweeps through the Brillouin zone (BZ) from $-\pi$ to $\pi$. \cite{Asboth}

Writing $\text{det}[Q(k)]=f(k)=R(k)e^{i\varphi(k)}$, it follows from the definition that $W=-(2\pi)^{-1}\int d\varphi$, or
\begin{equation} \label{Wdef}
W\,=\,-\frac{1}{2\pi}\int_{-\pi}^{\pi}d_k\varphi\,dk,
\end{equation}
where $d_k \equiv d/dk$. The minus sign in Eq. (\ref{Wdef}) is introduced to make $W>0$ since $\text{det}[Q(k)]$ winds clockwise, i.e. $d\varphi<0$.

Using that $i\varphi=\ln[f(k)]-\ln[R(k)]$, we get
\begin{equation}
W\,=\,-\frac{1}{2\pi i}\int_{-\pi}^{\pi}d_{k}\ln[f(k)]\,dk\,=\,-\frac{1}{2\pi i}\int_{-\pi}^{\pi}\frac{d_{k}f(k)}{f(k)}\,dk,
\end{equation}
since, differently from $f(k)$, $R(k)$ is a single-valued real function with $R(-\pi)=R(\pi)$.

The function $f$ depends on $k$ through $z=e^{-ik}$, i.e. $f=f(z(k))$ and thus
\begin{equation}
W\,=\,-\frac{1}{2\pi i}\int_{-\pi}^{\pi}\frac{d_{z}f(z(k))}{f(z(k))}d_{k}z(k)\,dk\,=\,\frac{1}{2\pi i}\oint\frac{d_{z}f(z)}{f(z)}\,dz,
\end{equation}
where the line integral is performed {\em counter-clockwise} around the unit circle in the complex plane.

Applying the Argument Principle of complex analysis, one finally arrives at
\begin{equation}\label{W2}
W\,=\,N_{z}-N_{p},
\end{equation}
where $N_{z}$ ($N_{p}$) is the number of zeros (poles) of $f(z)$ inside the unit circle, counting also their degrees (orders).

From Eqs. (\ref{SQk})-(\ref{SA}) one finds that, for any $r\in2\mathds{Z}$, $f(z)\,=\,\text{det}[Q(k)]\,=\,az^{2}+bz+c$ (and thus $\text{det}[Q(k)]$ indeed winds in the same clockwise direction as $z$), with $a,b,c$ constants depending on the parameters $\alpha_{n}^{\tau}$ and $\beta_{n}$. Therefore $N_{p}\,=\,0$ and $W\,=\,N_{z}\,=\,2,1,0$ if the zeros $z_{\pm}\,=\,(-b\pm\sqrt{b^{2}-4ac})/2a$ fall both inside, one inside and the other outside, or both outside the unit circle respectively.

Another useful formula to compute the winding number can be obtained by writing $\text{det}[Q(k)]=h_{x}(k)+ih_{y}(k)$ and thus $W=-(2\pi)^{-1}\int d\varphi$ becomes
\begin{equation}\label{W3}
W\,=\,\frac{1}{2\pi}\int_{BZ}\frac{h_{y}dh_{x}-h_{x}dh_{y}}{h_{x}^{2}+h_{y}^{2}},
\end{equation}
with $h_{x}$ and $h_{y}$ varying as $k$ sweeps through the BZ. \\

{\em Gapless surfaces.} \ There exist values of the microscopic parameters which enter $\alpha_{n}^{\tau}$ and $\beta_{n}$ (cf. Eqs. (\ref{alpha})-(\ref{beta})) - let us collectively call these values $\bar{g}$ - for which $\text{det}[Q(k)]=0$ at two points $k$ symmetrically located in the BZ - call them $k_{\pm}$. Due to Eq. (\ref{Hk}), also $\text{det}[{\cal H}(k_{\pm})]=0$, and hence ${\cal H}(k_{\pm})$ has at least one eigenvalue equal to zero. This fact, combined with chiral symmetry that forces the spectrum of ${\cal H}(k)$ to be symmetric around zero energy, implies that the gap between the bands immediately above and below zero energy closes at $k=k_{\pm}$. Moreover, since $\text{det}[Q(k_{\pm})]=0$, $W$ becomes undefined because the path traced out by $\text{det}[Q(k)]$ on the complex plane cuts exactly through the origin when $k=k_{\pm}$.

Following the proposal in Ref. \onlinecite{LYC}, when the spectrum is gapless we define ``local" winding numbers $\bar{W}_{\pm}$ computed \emph{around} each gap closing point $(\bar{g},k_{\pm})$. Specifically, $\bar{W}_{\pm}$ is defined as the number of times that $\text{det}[Q(k)]$ winds around the origin of the complex plane as the point $(g,k)$ moves, in the parameter-momentum space, around a small counter-clockwise contour $C_{\pm}$ centered at $(\bar{g},k_{\pm})$, with the latter being the single gap-closing point inside $C_{\pm}$. Analogous to the calculation of $W$ in Eq. (\ref{W3}), the invariants
$\bar{W}_{\pm}$ can thus be computed with the formula
\begin{equation}\label{Wlocal}
\bar{W}_{\pm}\,=\,\frac{1}{2\pi}\int_{C_{\pm}}\frac{h_{y}dh_{x}-h_{x}dh_{y}}{h_{x}^{2}+h_{y}^{2}},
\end{equation}
with $h_{x}$ and $h_{y}$ varying as $(g,k)$ winds around $C_{\pm}$.

Differently from $W$ which is a global invariant involving a computation over the entire BZ, $\bar{W}_{\pm}$ are local in the sense that they are defined in a small region in the parameter-momentum space. The \emph{accumulated} winding number characterizing a gapless spectrum is given by adding up the individual contributions from the two gap-closing points in the BZ: $\bar{W}=\bar{W}_{+}+\bar{W}_{-}$.
\begin{figure*}
\includegraphics[width=18cm]{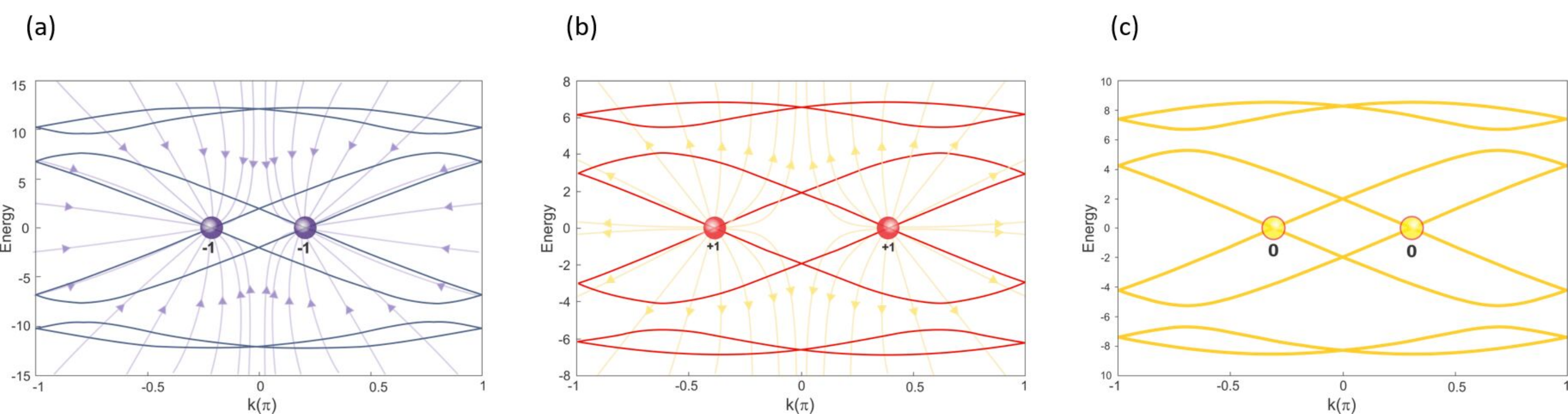}
\caption{(Color online) Spectrum for a point chosen on (a) the critical surface $A$ with $\bar{W}_{\pm}=-1$ ($\gamma_{\text{eff}}=6.0$, $\theta=0.50\pi$, $\phi=0.25\pi$), (b) the critical surface $B$ with $\bar{W}_{\pm}=1$ ($\gamma_{\text{eff}}=3.2$, $\theta=0.11\pi$, $\phi=0.30\pi$), (c) the intersection between $A$ and $B$ defining a multicritical line along which $\bar{W}_{\pm}=0$ ($\gamma_{\text{eff}}=4.1$, $\theta=0.88\pi$, $\phi=0.25\pi$). In (a)-(c), the zero-energy gap closes through the formation of a pair of time-reversal symmetric Dirac nodes. In (a) and (b) the nodes, which carry equal topological invariants $\bar{W}_{+}=\bar{W}_{-}$, are pictorially represented as repelling monopoles emanating fields that point towards or outwards from the source depending on the sign of the topological charge.}
\end{figure*}
\begin{figure*}
  \includegraphics[width=18cm]{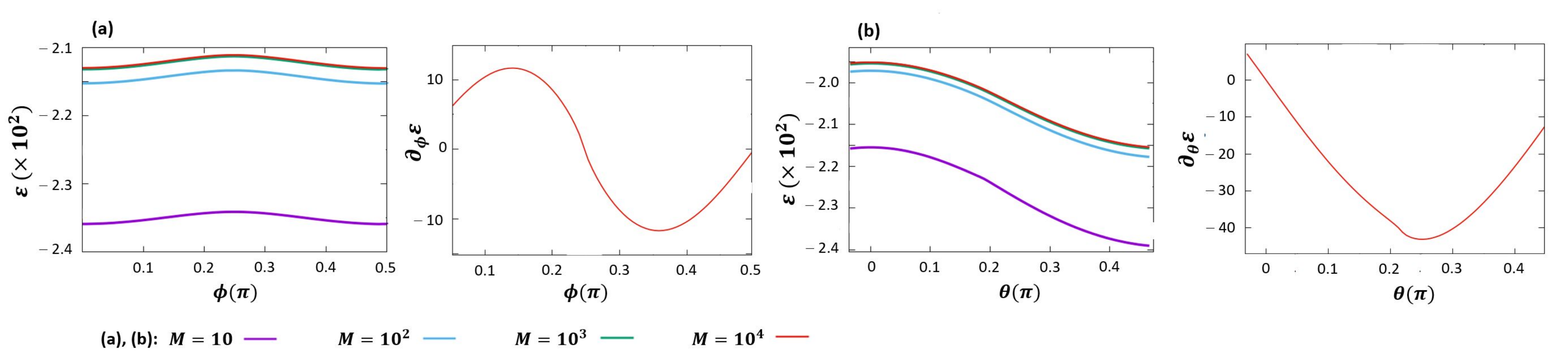}
  \caption{(Color online) (a), (b): Ground state energy density $\varepsilon$ (left panel) and its first derivative (right panel) for a topological QPT between a trivial $W\!=\!0$ phase and a topologically nontrivial $W\!=\!2$ phase across (a) the critical plane $A$ with $\theta=0.1\pi$ and (b) the critical surface $B$ with $\phi=0.1\pi$. In the plots, $M$ is the number of unit cells.}
\end{figure*}
\begin{figure*}
  \includegraphics[width=18cm]{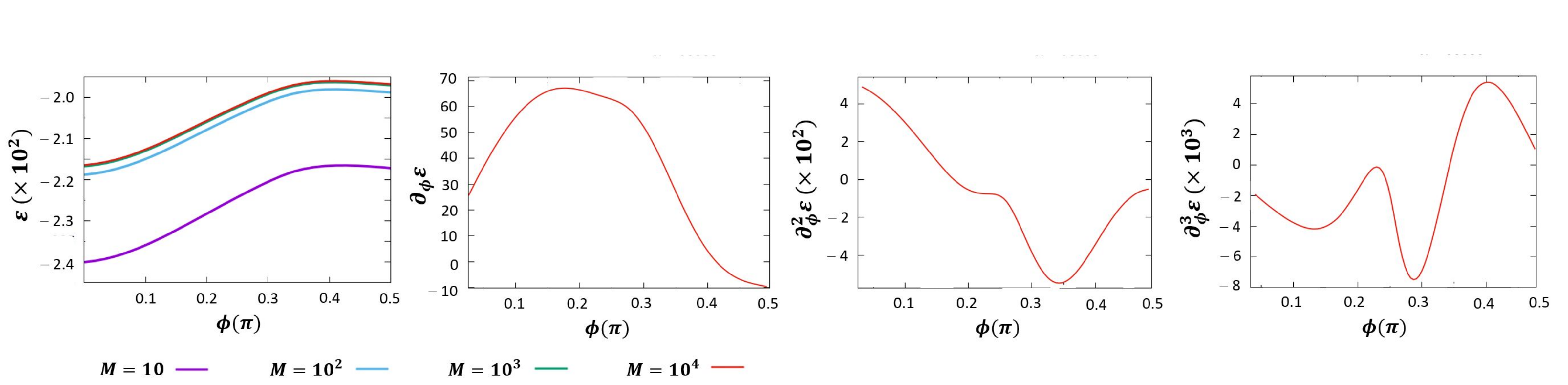}
  \caption{(Color online) Ground state energy density $\varepsilon$ and its derivatives up to third-order for a $W\,=\,0\,\rightarrow\,0$ path parametrized by $\theta=-(0.216/0.25)(\phi-0.5\pi)$ which cuts through the multicritical line. In the plots, $M$ is the number of unit cells.}
\end{figure*}

\subsection{Band structure and critical surfaces}

The critical (gapless) surfaces of the model come in two types: the plane $A$ at $\phi= \pi/4$, and the surfaces $B$ which are curved towards $\theta = 0, \pi$ for small $\gamma_{\text{eff}}$ and become flat for large $\gamma_{\text{eff}}$ (see Fig. 1). The intersections between the two types of surfaces define the multicritical lines which host the anomalous singularities identified in Sec. III.

When the model parameters $(\gamma_{\text{eff}}, \theta, \phi)$ take values on the plane $A$, the band gap in $k$-space closes at zero energy, resulting in two band degeneracies at points $k_{\pm}$ symmetrically located in the BZ (cf. the discussion above). Formally, by adopting an analysis from Ref. \onlinecite{Malard2018}, one finds that for any value of $\gamma_{\text{eff}}$ and $\theta$, choosing $\phi=\pi/4$,
\begin{equation}
{\cal M}(k) {\cal H}(k) {\cal M}^{-1}(k) = {\cal H}(-k)
\end{equation}
where ${\cal H}(k)$ is the Bloch matrix in Eqs. (\ref{Hk}), and ${\cal M}(k)$ is a matrix implementing a mirror transformation (composed by a site inversion and a spin flip). As follows from the analysis in Ref. \onlinecite{Malard2018}, the combination of this mirror $(M)$ symmetry with chiral $(S)$ and time-reversal $(T)$ symmetry enforces that any point on the plane $A$ supports two zero-energy and time-reversal symmetric band degeneracies in the BZ. In other words, these band degeneracies can be moved around in the BZ, but not removed, by changing a single parameter, $\gamma_{\text{eff}}$ or $\theta$.

As for the critical surfaces of type $B$, when the model parameters take values on these surfaces, the band gap also closes at zero energy, again resulting in zero-energy and time-reversal symmetric band degeneracies in the BZ. However, different from the degeneracies enforced by the $M, S,$ and $T$ symmetries on the plane $A$, these degeneracies are {\em accidental} because they are removable by changing the value of a single parameter, $\gamma_{\text{eff}}$ or $\theta$.

Three types of band structures with band degeneracies are shown in Fig. 5. Panel (a) ((b)) shows the spectrum for a point on a blue (orange) patch of the critical surface $A$ ($B$) in Fig. 1, i.e. where $\bar{W}_{\pm}=-1$ ($\bar{W}_{\pm}=1$). Fig. 5 (c) shows the spectrum for a point on a multicritical line where $A$ and $B$ intersect each other and where $\bar{W}_{\pm}=0$. In all cases, the gap closes at zero energy through the formation of a pair of time-reversal symmetric nodes, defining the apexes of two 1D Dirac cones and carrying equal winding numbers $\bar{W}_{+}=\bar{W}_{-}$. Figs. 5 (a)-(b) show a pictorial representation of the nodes as monopoles with associated fields.

\subsection{Ground state energy density and its derivatives}

Fig. 6 depicts the ground state energy density $\varepsilon$ for different values of the number $M$ of unit cells (left panel) and its smooth first derivative for $M=10^{4}$ (right panel) when going between a trivial $W\!=\!0$ phase and a topological $W\!=\!2$ phase across (a) plane $A$ and (b) surface $B$ (cf. Fig. 1). The left panels show that the thermodynamic limit - in which $\varepsilon$ is independent of $M$ - is achieved for $10^{3}<M<10^{4}$. For paths connecting two regions in the phase diagram with the {\em same} $W$, the derivatives of $\varepsilon$ are smooth up to third order, as seen in Fig. 7 for a $W\,=\,0\,\rightarrow\,0$ transition along the path $\theta=-(0.216/0.25)(\phi-0.5\pi)$ depicted in the inset of Fig. 1. By mirroring this path in the line $\phi=0.25\pi$ in the inset of Fig. 1, one realizes a $W\,=\,2\,\rightarrow\,2$ transition with first, second, and third energy derivatives obtained by mirroring the plots in Fig. 7 in the line $\phi=0.25\pi$. In both Figs. 6 and 7, $\gamma_{\text{eff}}=20$ (as in Fig. 2).

\end{document}